# Adaptive architecture towards portability of greenhouse models

L. Miranda[1,a] and G. Schillaci[2]

[1]Biosystems Engineering Group. Humboldt-Universität zu Berlin, Germany; [2]The BioRobotics Institute, Scuola Superiore Sant'Anna, Viale Rinaldo Piaggio, 34 - 56025 - Pontedera (PI) - Italy.

**Abstract**
   This work deals with the portability of greenhouse models, as we believe that this is a challenge to their practical usage in control strategies under production conditions. We address this task by means of adaptive neural networks, which re-adjust their weights when transferred to new conditions. Such an adaptive account for computational models is typical of the field of developmental robotics, which investigates learning of motor control in artificial systems inspired on infants development. Similarly to robots, greenhouses are complex systems comprising technical and biological elements, whose state can be measured and modified through control actions.
   We present an adaptive model architecture to perform online learning. This learning process makes use of an episodic memory and of online re-training, which modify the plasticity of the model according to the prediction error. This allows for adaptation without the need for a complete new training, which might be prohibitive if the data under the new conditions is scarce (in comparison to a research facility). Current experiments focus on how a model of tomato photosynthesis, developed in a research facility, can adapt itself to a new environment in a production greenhouse. Further research will focus on model plasticity by means of adaptive learning rates and management of the episodic memory described in this paper.
   The models presented as a proof-of-concept estimate the transpiration and photosynthesis of a hydroponic tomato crop by using measurements of the climate and the state of the actuators as inputs. In the thought experiment, the models are trained and tested using data from a greenhouse in Berlin, Germany. Thereafter, the adaptive architecture is fed with data from a production greenhouse in southern Germany, where other tomato varieties were grown under different irrigation and climate strategies. The proposed adaptive architecture represents a promising tool for spreading the use of models produced by high-tech research centers to the greenhouse production sector.

**Keywords:** Adaptive greenhouse models, online learning in deep neural networks, episodic memory, artificial intelligence, precision horticulture, photosynthesis, transpiration, tomato

**INTRODUCTION**
   This work presents an adaptive architecture for portability of greenhouse models. In particular, we propose adaptive neural networks which re-adjust their parameters when transferred to new conditions, by means of online learning processes and an episodic memory system.

   The proposed architecture deals with time series of multidimensional data, i.e.

---

[a] E-mail: mirandal@hu-berlin.de

measurements from different sensors and control actions in greenhouses. Learning the mappings between control actions and resulting measurements would allow the system to anticipate the effects of intervention to the greenhouse conditions and thus better plan further control actions. Furthermore, a capability to adapt (and therefore extrapolate model calculations) can facilitate the transfer of models from research facilities to the greenhouse industry.

As it will be described in this manuscript, state-of-the-art deep neural networks (a combination of Long Short-Term Memory recurrent neural networks, LSTM (Hochreiter and Schmidhuber, (1997) and non-recurrent neural layers) are adopted as machine learning tools to model short-term responses of a greenhouse crop to its environment. Specifically, this work focuses on portability of models between greenhouses and thus on the adaptivity of such LSTMs. In fact, the aim is to transfer a pre-trained model from a greenhouse condition, characterised by a specific mapping of the crop responses, to a different one, where the original mapping could not be consistent anymore. We investigate, therefore, the automatic online re-adjustment of the parameters of the neural network, so that the new mappings can be acquired as re-adaptation of the original ones, thus faster than performing a new training session with the new data from scratch.

However, readapting neural networks to map different input and output data than was not previously observed is not a trivial task. Simply re-fitting a pre-trained neural model with data recorded from a second environment is generally unfruitful, due to the limited extrapolation capabilities of neural networks. On the other hand, a completely new training of the network over newly-acquired data may lead to an issue known as catastrophic forgetting, where the information previously learned is quickly overwritten by the new incoming inputs (French, (1999)). Therefore, standard neural network learning techniques may not be feasible for portability of greenhouse models. We propose the usage of an episodic memory and of adaptive learning rates - described in the next sections - as a strategy to achieve this goal.

In the following section, we introduce relevant topics ranging from human brain development to cognitive and developmental robotics and describe how these concepts can be transferred into control system for greenhouses.

**GREENHOUSE MODEL PORTABILITY**

Depending upon their objectives and intentions, mathematical models can fall into two general forms: Mechanistic models and descriptive models. The former provide a degree of understanding or explanation of the subjacent processes taking place in the greenhouse and its subsystems, while descriptive models principally aim at describing the responses of the system as whole (without making any attempt to explain causal relationships) (Wallach et al. (2014), Thornley and France (2007)). When mechanistic models are used outside their original calibration domain, it is necessary to identify the concrete parameters that must be adjusted (e.g. greenhouse cover transmissivity to light). It is not possible to do the same with descriptive models, like neural models, because the model parameters are derived from data measurements, rather than theoretical relations. More empirical data would be needed to recalibrate these models before using them. Several authors have pointed out this drawback in the particular case of neural models of greenhouses.

Indeed, several studies have shown that neural networks are able to model different complicated processes occurring in a greenhouse, including internal climate (e.g. Linker et al. (1998); Uchida-Frausto and Pieters, (2004); Fitz-Rodríguez et al., (2012)), yield (e.g. Lin and Hill, (2008); Ehret et al., (2011); Salazar et al., (2015)) and physiological processes, such as transpiration (Zee and Bubenheim ,(1997)) and photosynthesis (Salazar-Moreno et al., (2011))

Their main problem, as reported in the literature, is their restricted extrapolation capability, which often bounds them to the system (greenhouse) from which the building data came (Seginer, 1997; Linker and Seginer, 2004). Seginer et al. (1994), after their research featuring neural networks trained in Avignon (France) and Silsoe (United Kingdom), also warn about using this kind of models outside their training domain.

**A DEVELOPMENTAL APPROACH TO GREENHOUSE MODEL PORTABILITY**

Adaptivity is one of the main characteristics of the human brain. Corporal dimensions and morphology change over time in humans and animals, as well as in plants. Several evidences suggest that the human brain maintains an internal representation of the body of the individual and that this representation undergo a continuous process of adaptation to cope with such a morphological development (Kaas, 1997; Cang and Feldheim, 2013; Haggard and Wolpert, 2005). In fact, neural pathways and synapses in the brain change with the behavior and the interaction of the individual with the environment, as well as with learning. Studies suggest that re-adaptations in the somatosensory cortex - the brain region which would maintain internal body representations - are triggered both by permanent (e.g. due to growth processes) and temporary (e.g. due to the usage of tools, see (Ganesh et al., 2014)) morphological changes.

During the last couple of decades, there has been a growing interest in the robotics community in the development of computational models inspired on the mechanisms of human brain adaptivity and on internal body representations. Recently, a new branch of the robotics research - i.e. developmental robotics - has gathered together scientists from different disciplines, including computer scientists, cognitive scientists, developmental psychologists and neuroscientists. The goal of developmental robotics is to produce autonomous, adaptive and social robots, which learn from and adapt to dynamic environments (Lungarella et al., 2003). The approach consists of applying to artificial systems models of human motor and cognitive development. Brain adaptivity, memory and curiosity-driven behaviours are, therefore, of great interest. Equipping robots with internal body representations, capable of adapting to dynamic circumstances and to the uncertainties of the human environment, would indeed improve their level of autonomy and interactivity (Schillaci et al., 2016). Similarly to humanoid robots, greenhouses are complex systems comprising technical and biological elements, whose state can be measured and modified through control actions. In this work, we propose a computational model that learns mappings between greenhouse sensory measurements and that re-adapts the acquired knowledge to different environmental conditions. In such systems, sensory information becomes progressively available and access to old information may be restricted, if not possible, for instance when a new greenhouse is being set up. The proposed learning mechanism makes use of an episodic memory, which prevents the disruptive interference that new training samples may have on the existing representations stored by the neural network.

**ADAPTIVE NEURAL ARCHITECTURE FOR GREENHOUSE MODELS**

In this work, we propose the adoption of LSTM recurrent neural networks. The developmental approach suggests to train these models in an incremental fashion, which requires the online update of the weights of the neural networks. This allows feeding the networks with new data, as soon as they are available. The advantage of taking this approach is multifaceted. First, this does not require the models to be pre-engineered for their new environment, as they are incrementally formed along the data that the system is faced to. Pre-defining complex systems a priori is, in fact, very challenging, if not impossible, as this would require exact knowledge about all the possible contingencies that can make the state of the system change and also about how the system will change due to these events. Second, the developmental approach allows the system to adapt to changing environmental conditions and thus enables portability of models.

As a proof of concept, we present an experiment on internal models trained on specific greenhouse conditions, which are thus re-adapted to new environmental and crop conditions. Our current efforts are also focused on extending the adaptive models to support developmental tools such as artificial curiosity to automatically trigger re-adaptation. In the next section, we present the tool of episodic memory, which we implemented in support to online learning processes.

**EPISODIC MEMORY**

Building adaptive neural networks is not a trivial task. A typical challenge is related to finding a proper balance between stability and plasticity of the computational models, which can ensure both long-term memory maintenance and propensity to sudden alteration of the system's conditions and of its surroundings.

Updating pre-trained neural networks to work on new tasks typically produces an issue known as catastrophic forgetting. In fact, the training on new samples may disrupt connection weights that were encoding previous mappings (Masse et al., 2018). Different strategies have been proposed to overcome this problem. Here, we propose the usage of an episodic memory. An incremental learning process produces online updates of the neural model, whenever new training samples are available. To reduce the catastrophic forgetting effects, an episodic memory system maintains a subset of previously experienced training samples and replays them, along with the new samples, to the networks during the training. The approach is known in the machine learning literature as system-level memory consolidation (McClelland et al.,1995), which takes inspiration from memory consolidation during sleep in humans. Evidences (e.g. Born and Wilhelm, 2012) suggest that sleep supports the formation of long-term memory, as experienced events are likely to be reactivated during sleep for consolidation. Memory consolidation is crucial especially in continual lifelong learning for artificial systems, where the aim is to program autonomous agents to interact in the real world and to progressively acquire, fine-tune and transfer knowledge over time spans (Parisi et al., 2019).

Moreover, a challenging task in continual lifelong learning is to find a good balance between how fast the system re-adapts to new knowledge and how long it maintains previously learned memories. This issue is known in the neurosciences - as well as in computer science and machine learning - as the stability-plasticity dilemma (Mermillod et al., 2013).

In the experiments presented here, we show the adaptive performance of a neural network encoding sensory data recorded from different greenhouses, by means of online learning and episodic memory. In the next section, we present the proof of concept. Future experiments will address the usage of adaptive learning rates and the analysis of the impact of different parametrisations of the system to the stability and plasticity of the model.

**PROOF OF CONCEPT**

In order to illustrate the implementation of a developmental model that uses an episodic memory to allow for greenhouse adaptation, we propose the following experiment.

Climate data (air temperature, relative humidity, solar radiation, $CO_2$ concentration) as well as leaf temperature were used as model inputs to an LSTM-network and used to predict the transpiration and photosynthesis rates of a tomato crop grown hydroponically under glass greenhouses at the Humboldt-Universität zu Berlin, in northern Germany. The available dataset comprised two similar greenhouses, with continuous (sampled every 5 minutes) measurements recorded from 2011 until 2016. The data from the first of these greenhouses (GH1) was fed to the models in an online fashion, that is, in an iterative loop that simulated an online streaming of data. More specifically, a batch of 100 data records

(each being a window of 250 subsequent measurements, the windows separated 10 minutes between them) was fed in each network iteration, corresponding to approximately one day of measurements. In other words, the model was incrementally trained each time with a batch of 100 samples.

Samples were progressively extracted from the recordings of cultivation years 2011 until 2014 of GH1, and passed, batch by batch, to a training scheme that updated the internal weights of the neural network to map the input-output contingencies. This process was implemented to simulate a continual and lifelong learning mechanism, where data is processed only when available. In order to tackle the catastrophic forgetting issue, during the online update, the model was presented not only with the latest observed samples, but also with a history of samples maintained in an episodic memory. In particular, the episodic memory contains a buffer of observed samples. At the beginning of the learning session, the buffer is empty and is filled up with the incoming sensory inputs-outputs. After having reached its full size (in this experiment, set to 10k samples), every time a new sample is observed, the content of the memory is updated as follows: iterate over all the elements of the memory and substitute it with the new observed samples with a probability of 0.1. This means, that the new sample can appear multiple times in the episodic memory. Our expectation is that this gives to the last observed sample a bigger impact in the model update (more plasticity), which would be otherwise trained using mostly previously observed samples and only the newly observed one (more stability). We plan to carry out further experiments on modulating the size of the memory and the probability of update and on analysing their impact on the stability and plasticity of the model, and therefore on its transferability.

After a first learning session on data from GH1, the same training process was applied onto data recorded from the second greenhouse (GH2), in particular sampled from the cultivation year 2015. Similarly, data from a third greenhouse (GH3), located in southern Germany (near Stuttgart), was used to further continue model training.

In order to track the learning progress of the model - that is, how well was the model encoded the relationship between input measurements, as specified above, and output measurements - we calculated a mean squared error (MSE) between the model predictions from test input measurements and ground truth output measurements. In particular, we extracted from each greenhouse (GH1, GH2 and GH3) a random subset of samples (10k samples for each greenhouse), each consisting of the input measurements, as specified above, and the corresponding output measurements. At a specific rate of online model updates (i.e. every three updates), we fed the input measurements of the test dataset to the neural network and compared the predicted outputs with the ground truth output measurements. Normalised input and outputs were used in both the training and testing phases. For the MSE calculations, test samples from the corresponding greenhouses have been used, while the model was being trained on each of the three greenhouse datasets.

Looking at the MSE values computed over time allows understanding whether there is a learning progress or not. A good learning curve should show an exponential diminishing trend, asymptotic to a steady-state error value. We expected to observe a decreasing MSE during online training on the first dataset (Greenhouse 1, 2011→2014), with sudden increase (or peaks) when presenting data from GH2 and GH3. The actual learning curves obtained are shown in Figure 1.

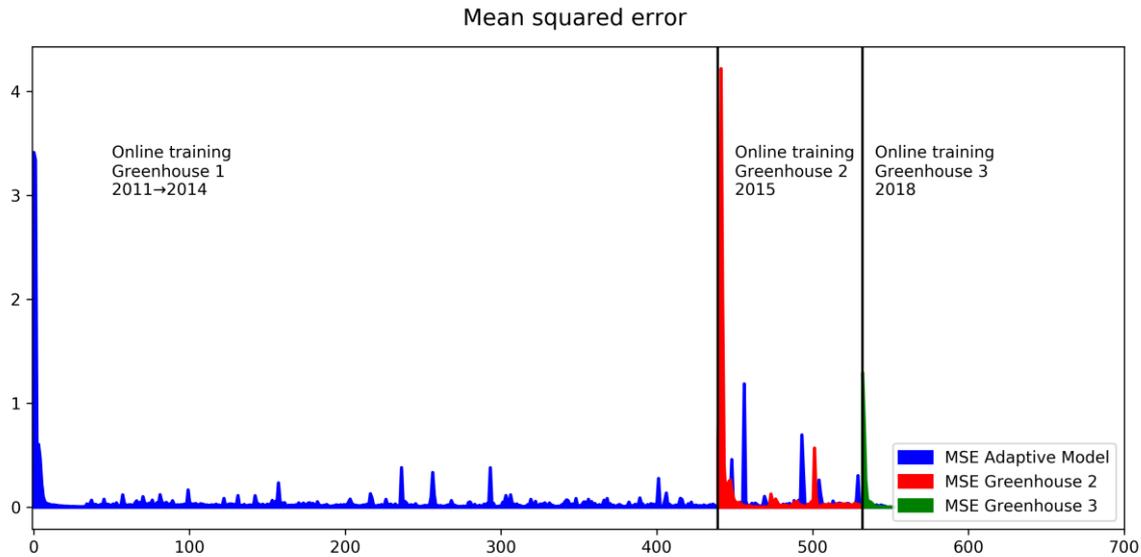

Figure 1. The trend of the MSE calculated on the test datasets during online learning. From left to right, the curves represent the error on a random validation set of data while online training in greenhouses 1, 2 and 3. The MSE is calculated for transpiration and photosynthesis normalized in [0:1].

In particular, the red curve shows the MSE calculated on GH1 test data, while the model is being updated in an online fashion with input-output samples extracted from GH1. After having processed all the samples from GH1 (around timestamp 430, corresponding to the first vertical line in the plot, from left to right), the algorithm starts presenting the model with samples extracted from GH2. From this point, the MSE is being calculated on GH2 test dataset. Similarly, after almost a hundred of MSE calculation steps (second vertical line in the plot), the system starts presenting the model with samples extracted from GH3 (and computing MSE values on GH3 test dataset). Interestingly, the transfer of the model to different environmental conditions did not impact negatively and significantly the learning progress. We explain this as being due to the usage of the episodic memory, which is maintaining many samples observed from the previous greenhouse, when switching to the new one. It has to be noted that GH1 and GH2 have similar environmental conditions (being both located in Berlin), compared to GH3 (located near Stuttgart). In fact, switching from GH1 to GH2 had a higher impact to the learning progress (blue peak in the MSE after the first vertical line). Nonetheless, the proposed approach performed successfully in the transfer of models between greenhouses. This is evident by comparing the performance of the model being trained online with data from GH1-GH2-GH3 (blue line in Figure 1), with the performance of other two models, each trained only on data from GH2 (red curve in the figure) and on data from GH3 (green curve in the figure), respectively. For the sake of convenience, the MSE curves of these models have been overimposed onto the blue curve, at the exact time when the first model is presented with data from GH2 and from GH3, respectively. The plots show that the initial performance of a brand new model is much worse (higher MSE values) than that of a model that has been transferred from another greenhouse. These results encourage spreading the use of similar models from high-tech research centers to the greenhouse production sector.

# CONCLUSIONS

The following conclusions can be drawn from the study. Online learning and adaptive neural network can enable the portability of greenhouse models. In particular, we proposed the usage of Long Short-Term Memory recurrent neural networks, updated in an online fashion. An episodic memory system has been proposed to tackle catastrophic forgetting issues in neural networks.

The implemented architecture showed very robust prediction capabilities when used in an unseen environment. The error levels achieved during online training in the first greenhouse were only marginally affected by migration to a second and third greenhouse, suggesting that this approach might be of use to transfer models between facilities. A particular case of interest would be the training of robust neural models in public research greenhouses and then making these models available for public usage. However, a number of research and engineering questions must be addressed before this point can be achieved.

We plan to carry out further experimentation to better understand the impact of the size of the episodic memory, the update rate of the memory and the learning rate of the model into the learning progress. Current efforts are also focused on modulating in real-time the adaptivity of the model, according to the trend of the MSE over time. Stability and plasticity, as previously discussed, can be modulated by re-configuring the memory size and the probability of update of its elements. Similarly, the learning rate, impacting the update of the weights of the neural network, can be modulated in real-time.

# ACKNOWLEDGEMENTS

The work of LM has partially received funding from the Project PROSIBOR. The project "Development of a sensor based intelligent greenhouse management system (PROSIBOR (FKZ: 2815701315)" is supported by funds of the Federal Ministry of Food and Agriculture (BMEL) based on a decision of the Parliament of the Federal Republic of Germany via the Federal Office for Agriculture and Food (BLE) under the innovation support programme.

The work of GS has received funding from the European Union's Horizon 2020 research and innovation programme under the Marie Sklodowska-Curie grant agreement No. 838861 (Predictive Robots) and under grant agreement No. 773875 (ROMI, Robotics for Microfarms).